\documentclass[12pt]{article}
\usepackage{amsmath,amssymb,amsthm}
\usepackage[T1]{fontenc}
\usepackage[utf8]{inputenc}
\usepackage[tt=false]{libertine}
\usepackage[scaled=0.83]{beramono}
\usepackage[libertine]{newtxmath}
\usepackage[dvips,dvipsnames,table]{xcolor}
\usepackage{graphicx}
\usepackage{enumitem}
\usepackage{bm}
\usepackage{placeins}
\usepackage{authblk}
\usepackage[margin=1in]{geometry}
\usepackage{setspace}
\usepackage[backref, natbib ,style = nature ,backend = bibtex, sorting=none]{biblatex}
\definecolor{ptCol1}{HTML}{18453b}
\definecolor{ptCol2}{HTML}{d4895e}
\definecolor{ptCol3}{HTML}{825d83}
\definecolor{ptCol4}{HTML}{dfad5d}
\definecolor{ptCol5}{HTML}{506275}
\definecolor{ptCol6}{HTML}{00585a}
\definecolor{ptCol7}{HTML}{aa4222}
\definecolor{ptCol8}{HTML}{a8b9c9}
\definecolor{ptCol9}{HTML}{386930}
\definecolor{MSUGreen}{HTML}{18453b}
\definecolor{ptOrange}{HTML}{d4895e}
\definecolor{ptPurple}{HTML}{825d83}
\definecolor{ptYellow}{HTML}{dfad5d}
\definecolor{ptBlue}{HTML}{506275}
\definecolor{ptTurquoise}{HTML}{00585a}
\definecolor{ptRed}{HTML}{aa4222}
\definecolor{ptLightBlue}{HTML}{a8b9c9}
\definecolor{ptGreen}{HTML}{386930}
\definecolor{tablerows}{HTML}{E5EBE7}
\definecolor{tablerowsLine}{HTML}{ABBFB2}

\usepackage{caption}
\usepackage{subcaption}
\newlength{\mycaptionmargin}
\newlength{\mycaptionwidth}
\DeclareCaptionLabelFormat{underlcap}{\underline{#1 #2}}
\usepackage{cleveref}
\crefname{figure}{figure}{figures}
\crefname{table}{table}{tables}
\newcommand{\initem}[1]{\textcolor{MSUGreen}{(#1)}}

\newcommand{\tskip}[1]{\vphantom{\rule{0pt}{#1em}}}
\definecolor{tablerows}{HTML}{E5EBE7}

\begin{document}

\title{From Genotype to Phenotype: polygenic prediction of complex human traits}

\author[1]{Timothy G. Raben}
\author[1,2]{Louis Lello}
\author[1]{Erik Widen}
\author[1,2]{Stephen D.H. Hsu}
\affil[1]{Michigan State University, East Lansing, Michigan, 48824}
\affil[2]{Genomic Prediction, North Brunswick, New Jersey, 08902}
\date{}

\maketitle

\begin{abstract}
Decoding the genome confers the capability to predict characteristics of the organism (phenotype) from DNA (genotype). We describe the present status and future prospects of genomic prediction of complex traits in humans. Some highly heritable complex phenotypes such as height and other quantitative traits can already be predicted with reasonable accuracy from DNA alone. For many diseases, including important common conditions such as coronary artery disease, breast cancer, type I and II diabetes, individuals with outlier polygenic scores (e.g., top few percent) have been shown to have 5 or even 10 times higher risk than average. Several psychiatric conditions such as schizophrenia and autism also fall into this category. We discuss related topics such as the genetic architecture of complex traits, sibling validation of polygenic scores, and applications to adult health, in vitro fertilization (embryo selection), and genetic engineering.
\end{abstract}

\noindent  A version of this article was prepared for \textbf{Genomic Prediction of Complex Traits}, Springer Nature book series \textbf{Methods in Molecular Biology}.

\bigskip

\noindent\textbf{Keywords:} genomics, complex trait prediction, PRS, in vitro fertilization, genetic engineering

\section{Introduction}

\noindent {\it I, on the other hand, knew nothing, except ... physics and mathematics and an ability to turn my hand to new things.} ― Francis Crick
\smallskip

The challenge of decoding the genome has loomed large over biology since the time of Watson and Crick. Initially, decoding referred to the relationship between DNA and specific proteins or molecular mechanisms, but the ultimate goal is to deduce the relationship between DNA and phenotype --- the character of the organism itself. {\it How does Nature encode the traits of the organism in DNA?} $\,$ In this review we describe recent advances toward this goal, which have resulted from the application of machine learning (ML) to large genomic data sets. Genomic prediction is the real decoding of the genome: the creation of mathematical models which map genotypes to complex traits.

It is a peculiarity of ML and artificial intelligence (AI) applied to complex systems that these methods can often ``solve'' a problem without explicating, in a manner that humans can absorb, the intricate mechanisms that lie intermediate between input and output. For example, AlphaGo \cite{gibney2016google} achieved superhuman mastery of an ancient game that had been under serious study for thousands of years. Yet nowhere in the resulting neural network with millions of connection strengths is there a human-comprehensible guide to Go strategy or game dynamics. Similarly, genomic prediction has produced mathematical functions which predict quantitative human traits with surprising accuracy --- e.g., height, bone density, and cholesterol or lipoprotein A levels in blood (see Table \ref{tab:quantTraits}); using typically thousands of genetic variants as input (see next section for details) --- but without explicitly revealing the role of these variants in actual biochemical mechanisms. Characterizing these mechanisms --- which are involved in phenomena such as bone growth, lipid metabolism, hormonal regulation, protein interactions --- will be a project which takes much longer to complete. 

\begin{table}%{l}{300 pt}
 \newcommand{\pcell}[2]{\makecell[l]{#1\\[-0.5ex]~~#2}}
\centering
\rowcolors{2}{}{tablerows}
\small
\begin{tabular}{l l l}
      Phenotype & Correlation & \# active SNPs \\\hline
\tskip{1.2}Height						 & $0.639_{(0.017)}$ 	& $22,000_{(3000)}$ 				\\
Heel bone density				 & $0.449_{(0.015)}$ 	& $15,000_{(4000)}$				\\
BMI						 & $0.346_{(0.0009}$	& $22,000_{(2000)}$	\\
Educational attainment				 & $0.272_{(0.022)}$	& $17,000_{(7000)}$			\\
Apolipoprotein A                                 & $0.417_{(0.006)}$	& $15,000_{(2,000)}$	 \\
Apolipoprotein B                                 & $0.38_{(0.01)}$	& $9,000_{(2,000)}$	 \\
Cholesterol                                      & $0.310_{(0.007)}$	& $10,000_{(3,000)}$	 \\
Direct bilirubin                                 & $0.51_{(0.01)}$	& $4,000_{(4,000)}$	 \\
HDL cholesterol                                  & $0.46_{(0.01)}$	& $17,000_{(2,000)}$	 \\
Lipoprotein A                                    & $0.757_{(0.008)}$	& $3,000_{(1,000)}$	 \\
Platelet count                                   & $0.45_{(0.01)}$	& $15,000_{(800)}$	 \\
Total bilirubin                                  & $0.56_{(0.01)}$	& $5,000_{(3,000)}$	 \\
Total protein                                    & $0.32_{(0.01)}$	& $15,000_{(1,000)}$	 \\
Triglycerides                                    & $0.348_{(0.008)}$	& $11,000_{(4,000)}$	 \\
\end{tabular}
\caption{Examples of quantitative trait prediction. The last 10 traits listed are all obtained from standard blood test measurements (terminology from UK Biobank data fields). Uncertainties given in parenthesis are the standard deviation obtained from validation of 5 different training runs, each of which produce a slightly different predictor. Predictors were trained with data and methods analogous to \cite{predictors}.}
\label{tab:quantTraits}
\end{table}

\begin{figure}[h!]
\includegraphics[width=0.95\linewidth]{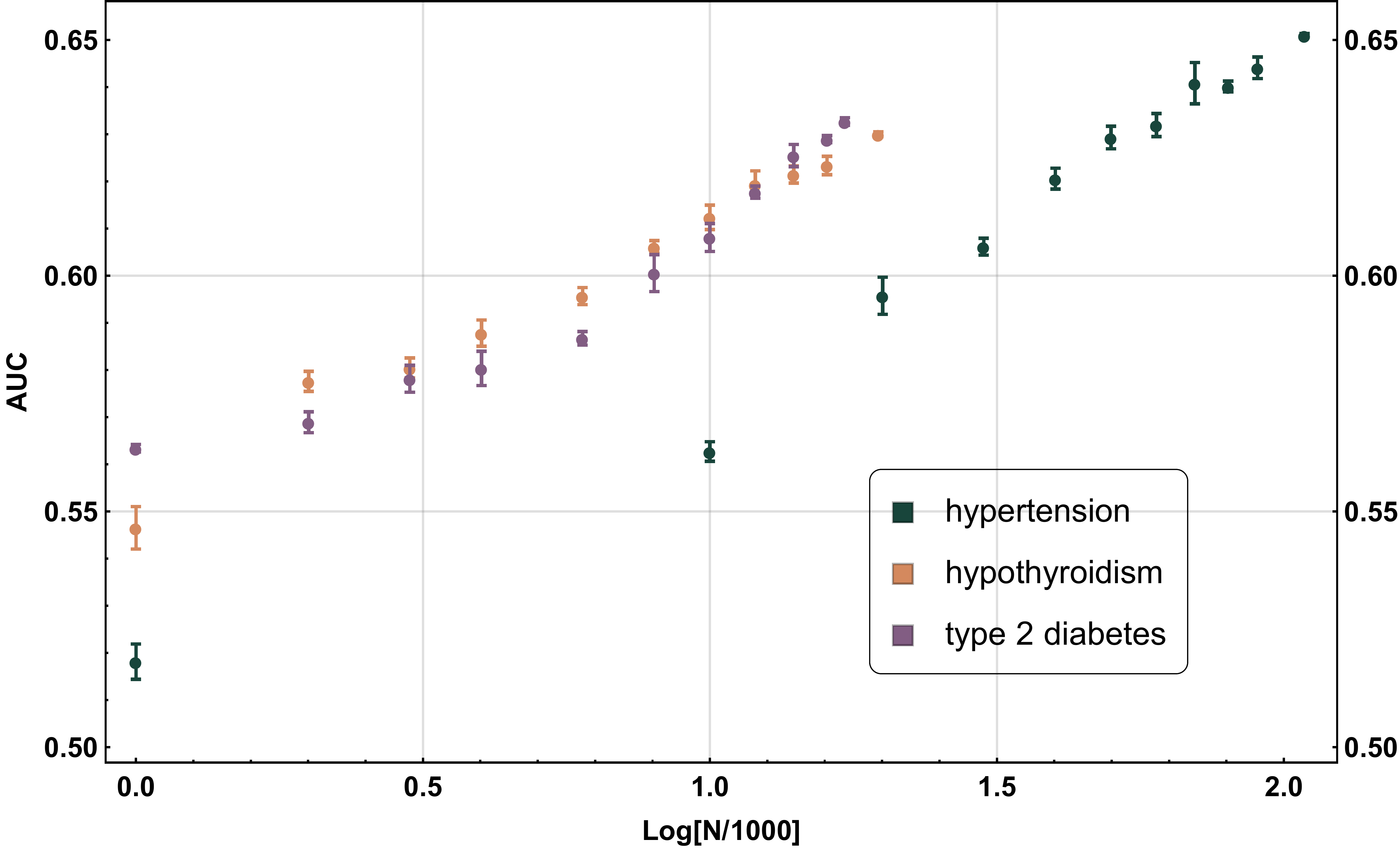}
\caption{
Prediction quality measured by AUC for three conditions (hypertension, hypothyroidism, type 2 diabetes) as a function of training sample size $N$. Evidence is strong that further increases in sample size will lead to improvement in accuracy. Reproduced from \cite{predictors}.
}
\label{fig:auc}
\end{figure}

If recent trends persist, in particular the continued growth of large genotype | phenotype data sets, we will likely have good genomic predictors for a host of human traits within the next decade. Here {\it good} can mean capturing {\it most} of the a priori estimated heritability of the trait. There are already many disease risk predictors which fall somewhat short of this standard but nevertheless have important practical utility in medicine (e.g., for early screening and diagnosis), as we will discuss below. We can roughly estimate how predictor performance increases as a function of training data size (see \cref{fig:auc}). The estimates suggest that progress is {\it data limited}: algorithms and computational resources are not the bottleneck.

In the following sections we will discuss \initem{1} Current status of human genomic prediction: examples of quantitative trait prediction and prediction of disease risk, out of sample validation of predictors, sibling validation, \initem{2} Methods, Genetic Architecture, and Theory, and \initem{3} Applications to In Vitro Fertilization and Gene Editing. The final section will discuss some (near term) future projections.

\section{Present Status: 2020}

Technological advances have reduced the (2020) cost of whole genome sequencing to under \$1k and the cost of SNP array genotyping to roughly \$20 \cite{genomecost}. In this section we will give an overview of results obtained from training on data sets obtained using SNP arrays. For quantitative traits sample sizes were in excess of 400k individuals. For disease risk training (polygenic risk scores, or PRS) typical sample sizes were tens of thousands of cases and at least as many controls, typically individuals late in life for whom medical records are available. See \cite{predictors} for specific details.

\subsection{Quantitative traits: height, bone, and blood}
\begin{figure}[h!]
\includegraphics[width=0.95\linewidth]{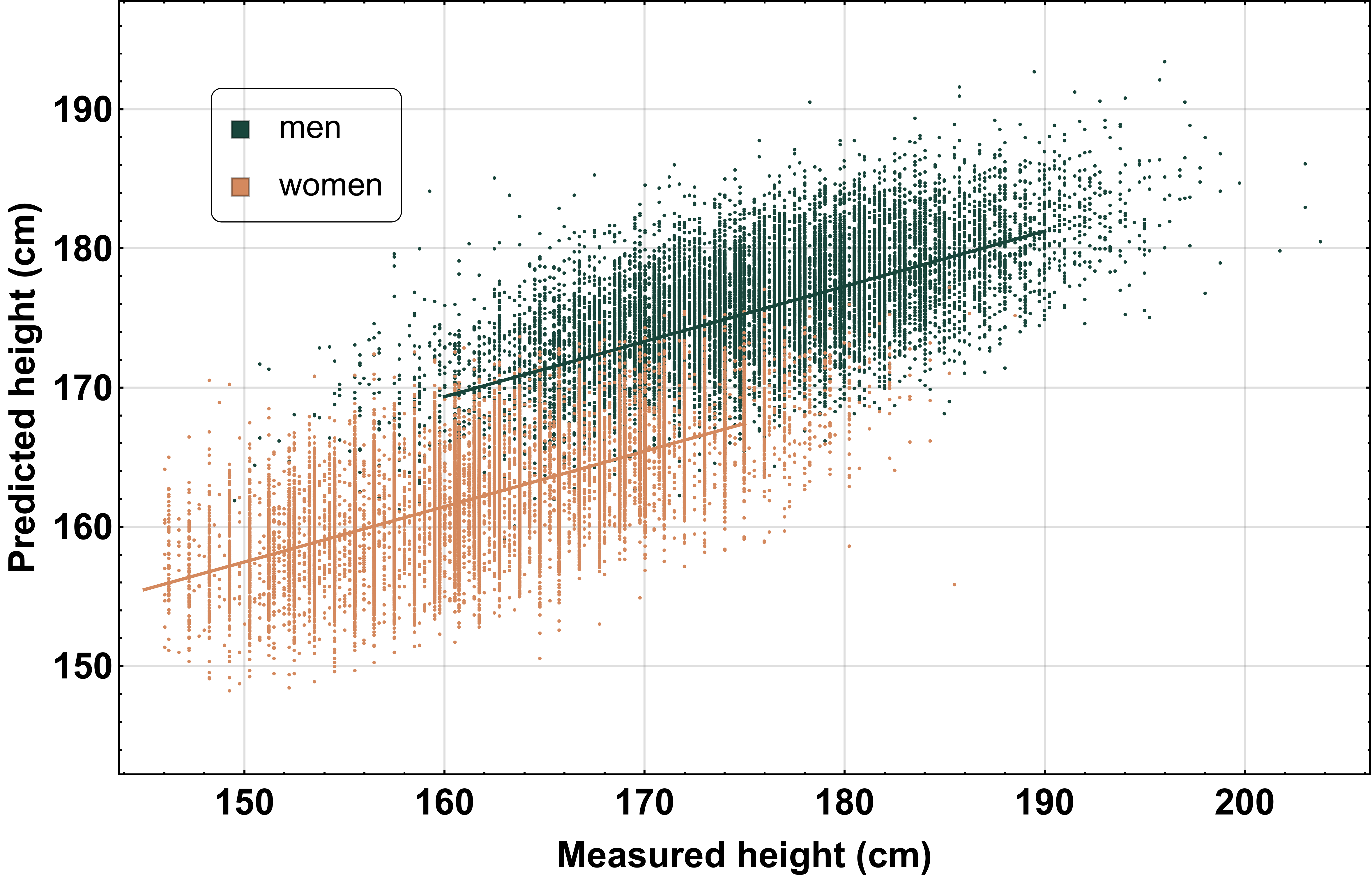}
\caption{
Height prediction in males and females not used in predictor training. Predicted height computed from $\sim$20k SNPs is shown on vertical axis, and actual height on horizontal axis. In a typical bin the standard deviation of the distribution of actual heights relative to prediction is $\sim 3$cm. While it is possible to find individuals whose height deviates from prediction by significantly more than a few cm, most of the population density is concentrated close to the prediction line. Correlation between predicted and actual (z-scored) height is $\sim 0.65$. Reproduced from \cite{height}.
}
\label{fig:height}
\end{figure}

In 2012 one of the authors \cite{Ho2015} estimated that training on a few hundred thousand SNP genotypes using L1 penalization (see next section for details) would capture most of the common SNP heritability for height. In 2017, with the release of the UK Biobank data set of ~500k genomes, this prediction came true. Accurate genomic prediction of adult height, with standard deviation of $\sim 3$ cm, was demonstrated in \cite{height}, and has been subsequently replicated by other groups \cite{Chung2019,Qian2020}, as well as in studies of siblings. See \cref{fig:height,fig:heightfrac} for a demonstration of the prediction accuracy.

In 2020 the GIANT collaboration, in a GWAS of roughly 4 million individuals, identified $\sim$10k height SNPs at genome-wide statistical significance. The variance accounted for by the L1 predictor and by the predictor produced from the 2020 GIANT GWAS are roughly equal to the previously estimated narrow sense heritability accounted for by common SNPs \cite{Yengo2020,Kaiser2020}. 

Genomic predictors which capture a significant fraction of heritability exist for other quantitative traits, including bone density, educational attainment / cognitive abilities, and a number of blood measurements such as platelet count and lipoprotein A levels, see \cref{tab:quantTraits}.

\subsection{Disease Risks: Polygenic Risk Scores (PRS)}
Polygenic risk predictors for dozens of important disease conditions, including, e.g., diabetes, breast cancer, coronary artery disease, hypertension, schizophrenia, autism, and many more, have been published and validated by many research groups \cite{predictors,khera,khera2019polygenic,LewisVassos202005}.

We can roughly characterize the performance of these polygenic risk predictors as follows: individuals with very high PRS will typically have an incidence rate which is many times higher than the population average. For example, in \cite{predictors} we found that for atrial fibrillation, 99th percentile PRS implies $\sim$10 times higher likelihood of case status. Similarly, low PRS indicates below average risk for the condition: we can identify individuals whose risk is an order of magnitude lower than in the general population. 

\begin{figure}[h!]
\includegraphics[width=0.45\linewidth]{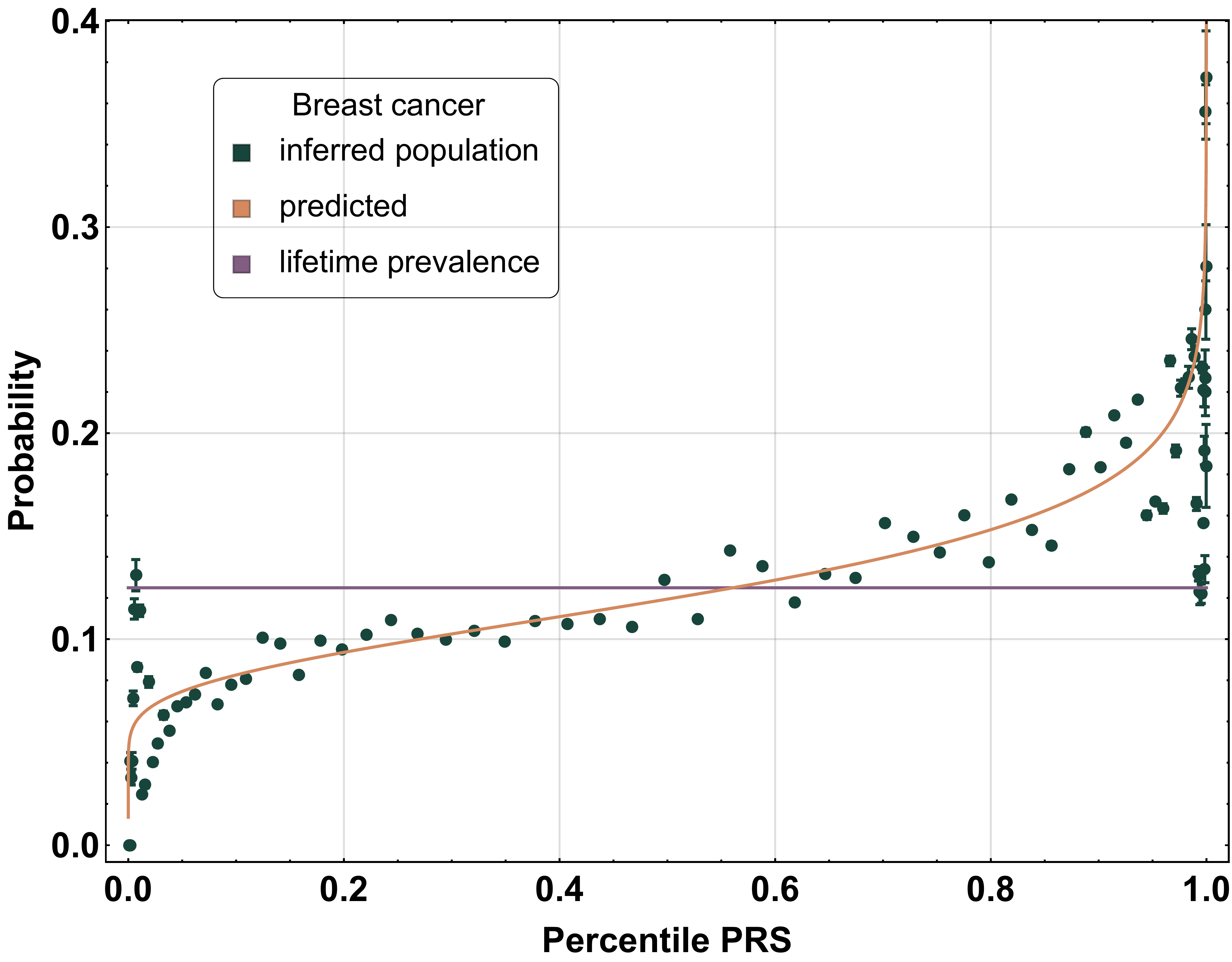}
\includegraphics[width=0.45\linewidth]{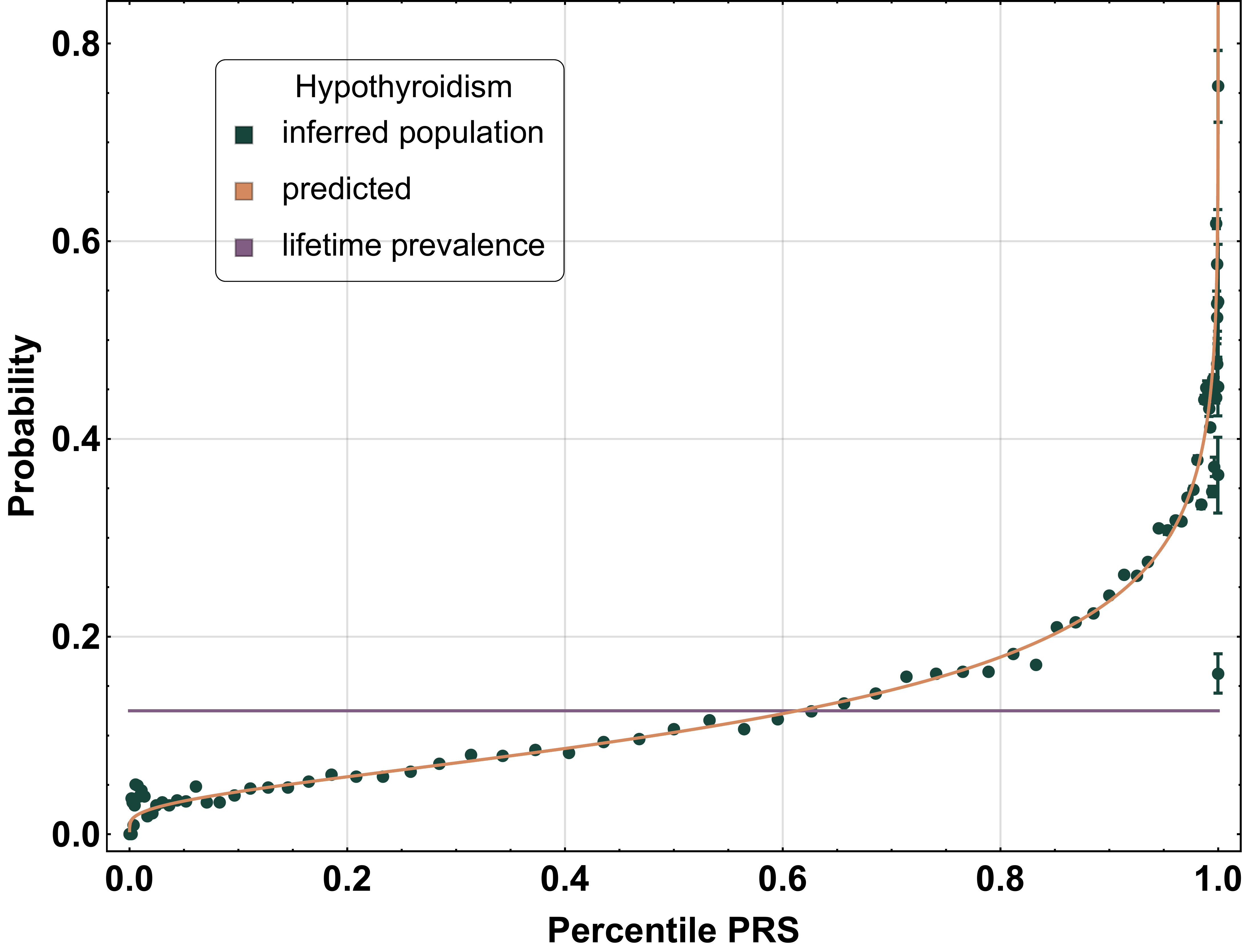} \\
\caption{
Incidence of breast cancer and hypothyroidism as a function of percentile polygenic risk score (PRS). At high PRS the likelihood of incidence increases nonlinearly, and at low PRS the likelihood decreases nonlinearly. Red curve is theoretical, modeling case and control populations with normal distributions shifted in mean PRS. Blue data points are calculated using individuals (not used in training) binned by PRS. Reproduced from \cite{predictors}.
}
\label{fig:prevalence}
\end{figure}

Identification of outliers is possible even though the standard performance metric AUC (Area Under ROC Curve) value is modest: e.g., AUC $\sim 0.6$. This is because the absolute risk as function of PRS is highly nonlinear: outlier (e.g., 99th percentile) risk can be very high even if risk for individuals near the middle of the PRS distribution varies only modestly --- see \cref{fig:prevalence}, which illustrates risk differentiation for the example phenotypes breast cancer and hypothyroidism. We report below on further explicit results.

\begin{figure}[h!]
\includegraphics[width=0.45\linewidth]{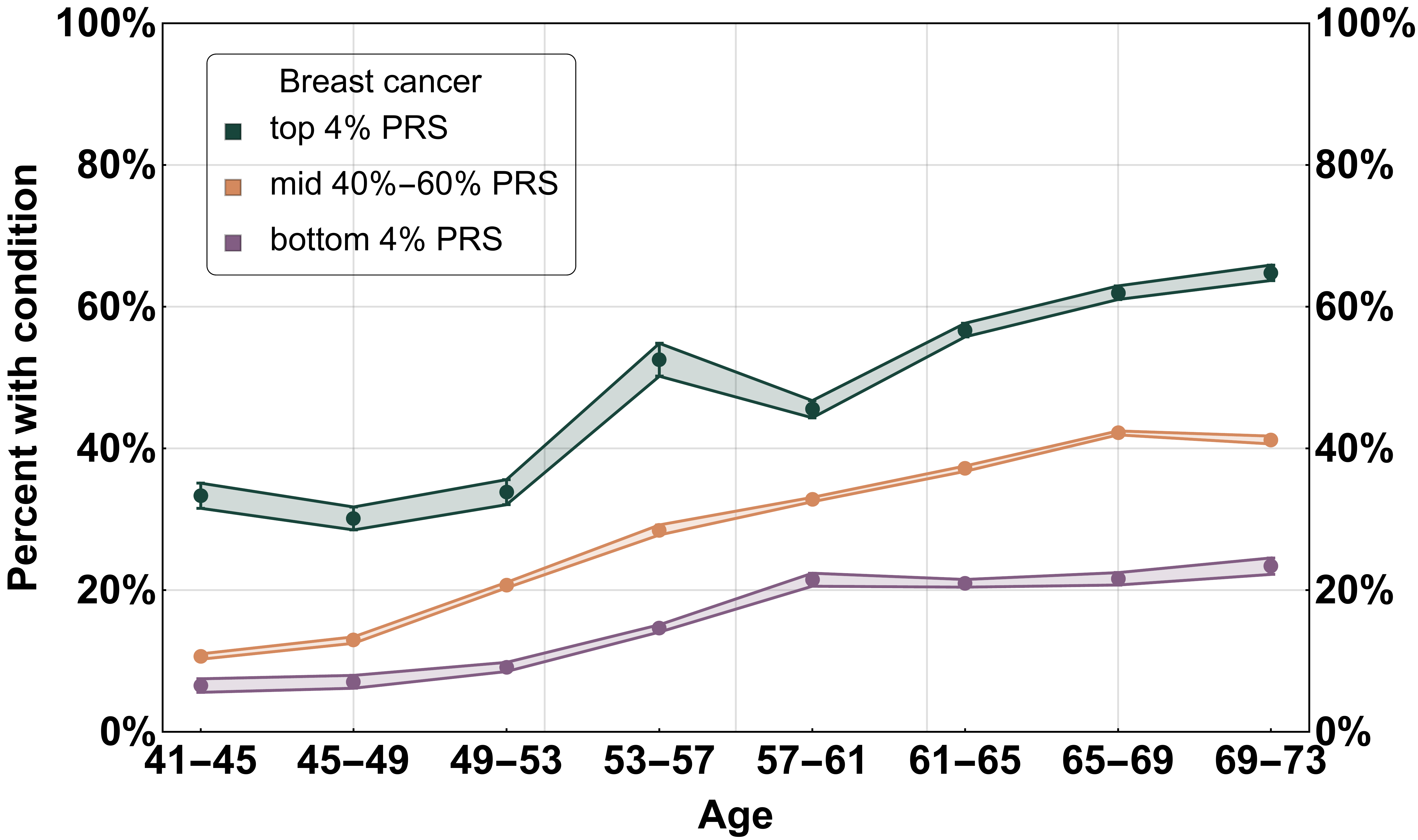}
\includegraphics[width=0.45\linewidth]{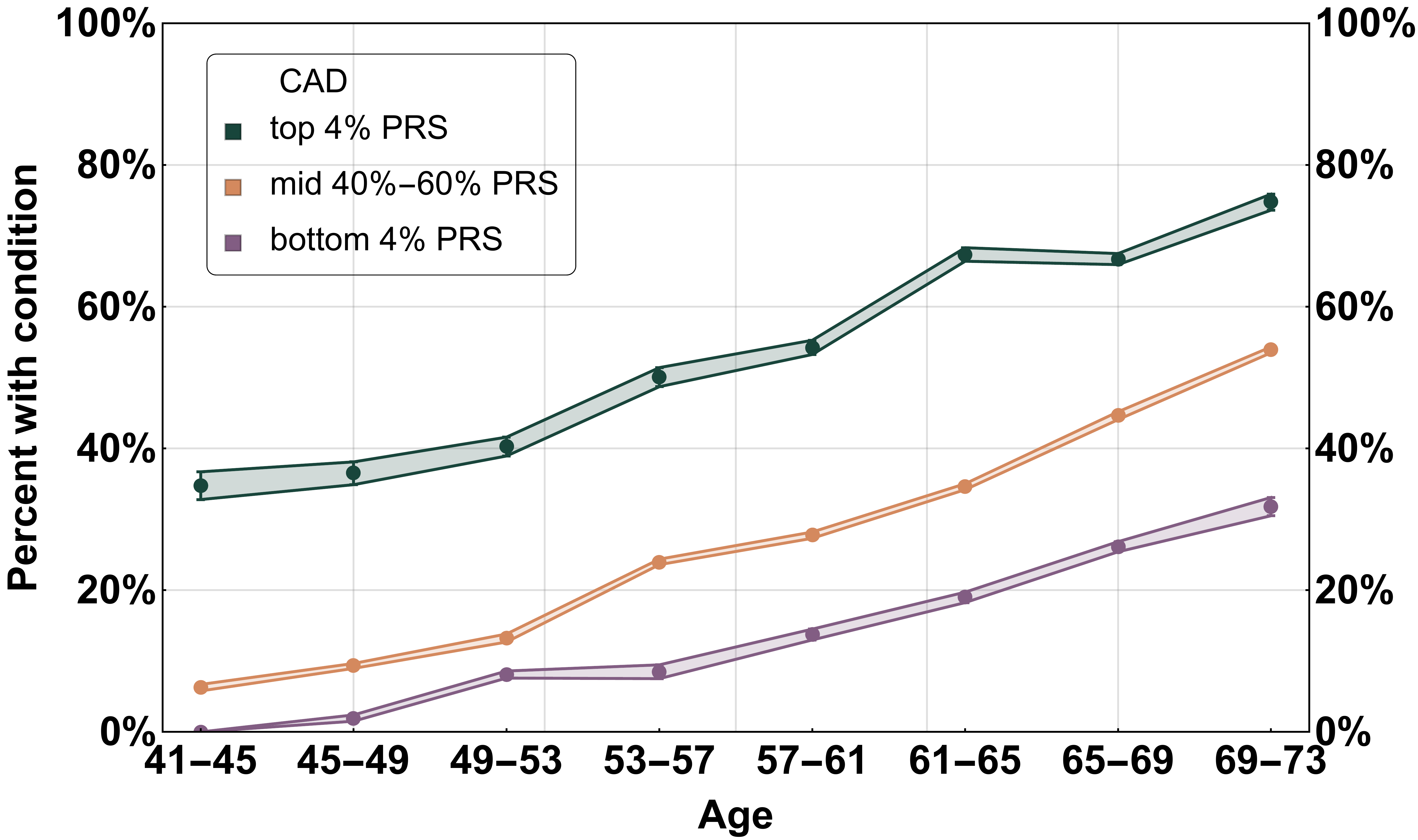} \\
\caption{
Breast cancer and coronary artery disease (CAD) risk progression with age and polygenic score (PRS). Outlier individuals with unusually high (top 4 percent) PRS are much more likely to be diagnosed with the condition than those with typical (40th to 60th percentile) PRS. Low PRS is associated with reduced incidence of the condition. Similar results are available for 20 or more common disease conditions, and have obvious utility for early screening, diagnosis, and prevention. PRS obtained from \cite{khera} and scored on a population within the UK Biobank \cite{ukbb-nature}.}
\label{fig:agerisk}
\end{figure}

There is already significant interest in the application of PRS in a clinical setting, for example to identify high risk individuals who might receive early screening or preventative care \cite{torkamani2018personal,liu2018genome,chatterjee2016developing,euesden2014prsice,clinicalPRS,clinicalbreast,predictors,clinicalprospects,clinicalapplication,moreclinical1,moreclinical2,moreclinical3,moreclinical4}. As a concrete example, women with high PRS scores for breast cancer can be offered early screening: already standard of care for those with BRCA risk variants \cite{breastrisk1,breastrisk2}. However, BRCA mutations affect no more than a few women per thousand in the general population \cite{BRCAfrequency1,BRCAfrequency2,BRCAfrequency3}. Importantly,  the number of (BRCA negative) women who are at high risk for breast cancer due to polygenic effects is an order of magnitude larger than the population of BRCA carriers  \cite{khera,predictors,PRSrisk,prsbreast1,prsbreast2,myriad2020-08,myriadWebsite}. From this one example it is clear that significant medical, public health, and cost benefits could result from PRS (e.g. \cite{breastcost}). It is well known that patients with atherosclerotic diseases, coronary artery disease (CAD), and lung diseases can benefit from early intervention \cite{farpour2009physical,mehta2009early,busse2008inhaled}.  In many instances where early treatment can be beneficial, stratification by age, gender, and ethnicity show an exacerbation of poor outcomes \cite{bhalotra2007disparities}. Precision genetics is already used in identification of candidates for early intervention, and will become widespread in the near future (cf. Myriad's riskScore test and other examples \cite{myriadWebsite,myriad2020-08}). In \cref{fig:agerisk}, we illustrate the predicted risk of breast cancer and coronary artery disease as function of age for high, medium and low risk groups, respectively.

It should be stressed that here we focus on \emph{purely genomic} risk scores and correlations. That is, we are focused on relative genetic risk from SNP information alone. These results can be easily combined with information from other biomarkers (e.g., blood test results) or health-related indicators such as age, sex, BMI, blood pressure, etc. to obtain even stronger risk stratification \cite{predictors,khera,khera2019polygenic,chatterjee2016developing}.

PRS with similar utility for risk outlier identification have been developed for psychiatric conditions such as autism, schizophrenia \cite{Grove2019,Ripke2014}.

\subsection{Sibling Validation}
There are now many validations of polygenic prediction in the scientific literature, conducted using groups of people born on different continents and in different decades than the original populations used in training \cite{Wunnemann2019, belsky2018genetic}. Here we discuss results showing that predictors can differentiate between siblings (which one has heart disease? is taller?), despite similarity in childhood environments and genotype. The predictors work almost as well in pairwise sibling comparisons as in comparisons between randomly selected strangers. 

\begin{figure}[h!]
\includegraphics[width=0.45\linewidth]{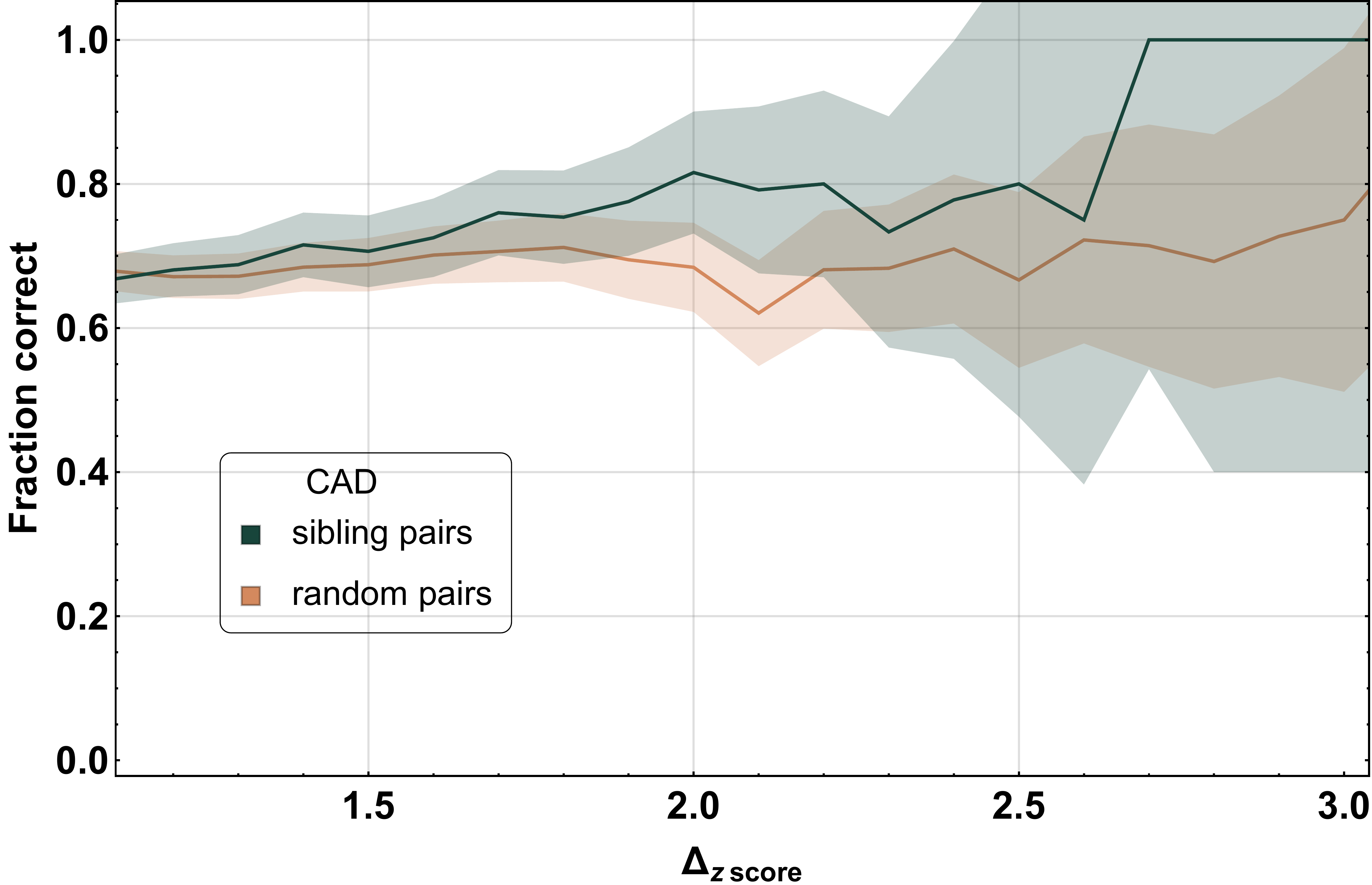}
\includegraphics[width=0.45\linewidth]{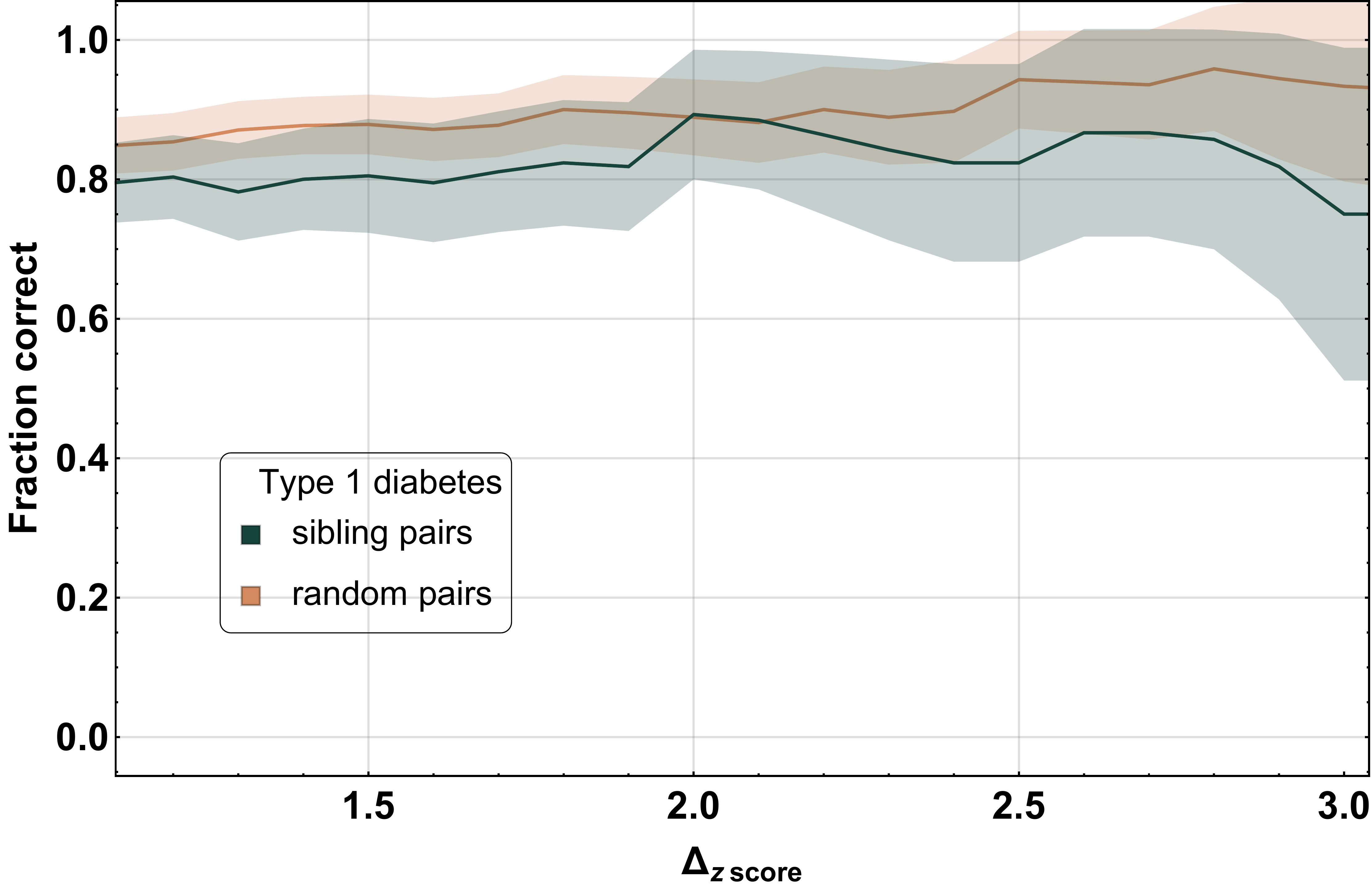} \\
\includegraphics[width=0.45\linewidth]{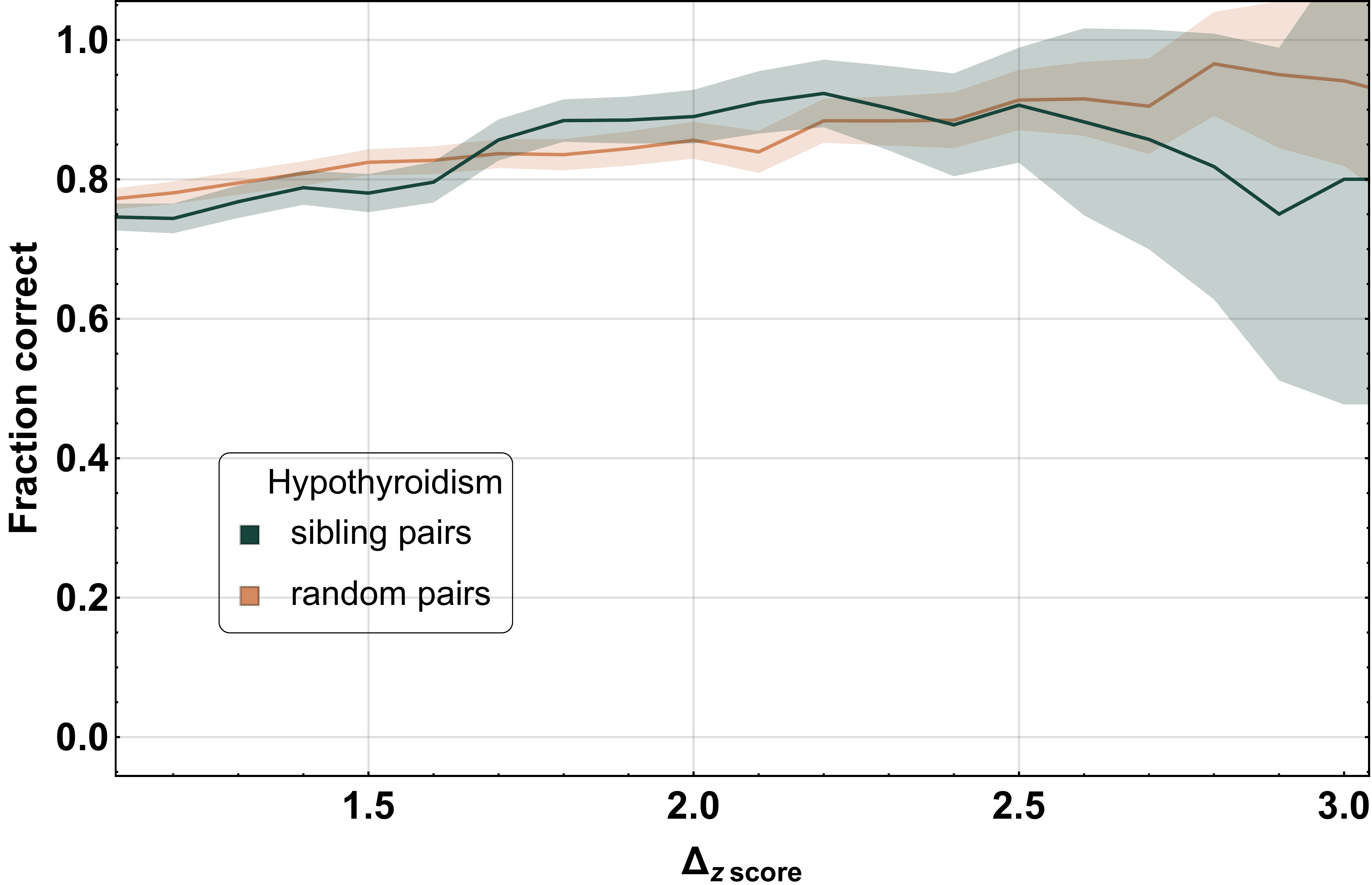}
\includegraphics[width=0.45\linewidth]{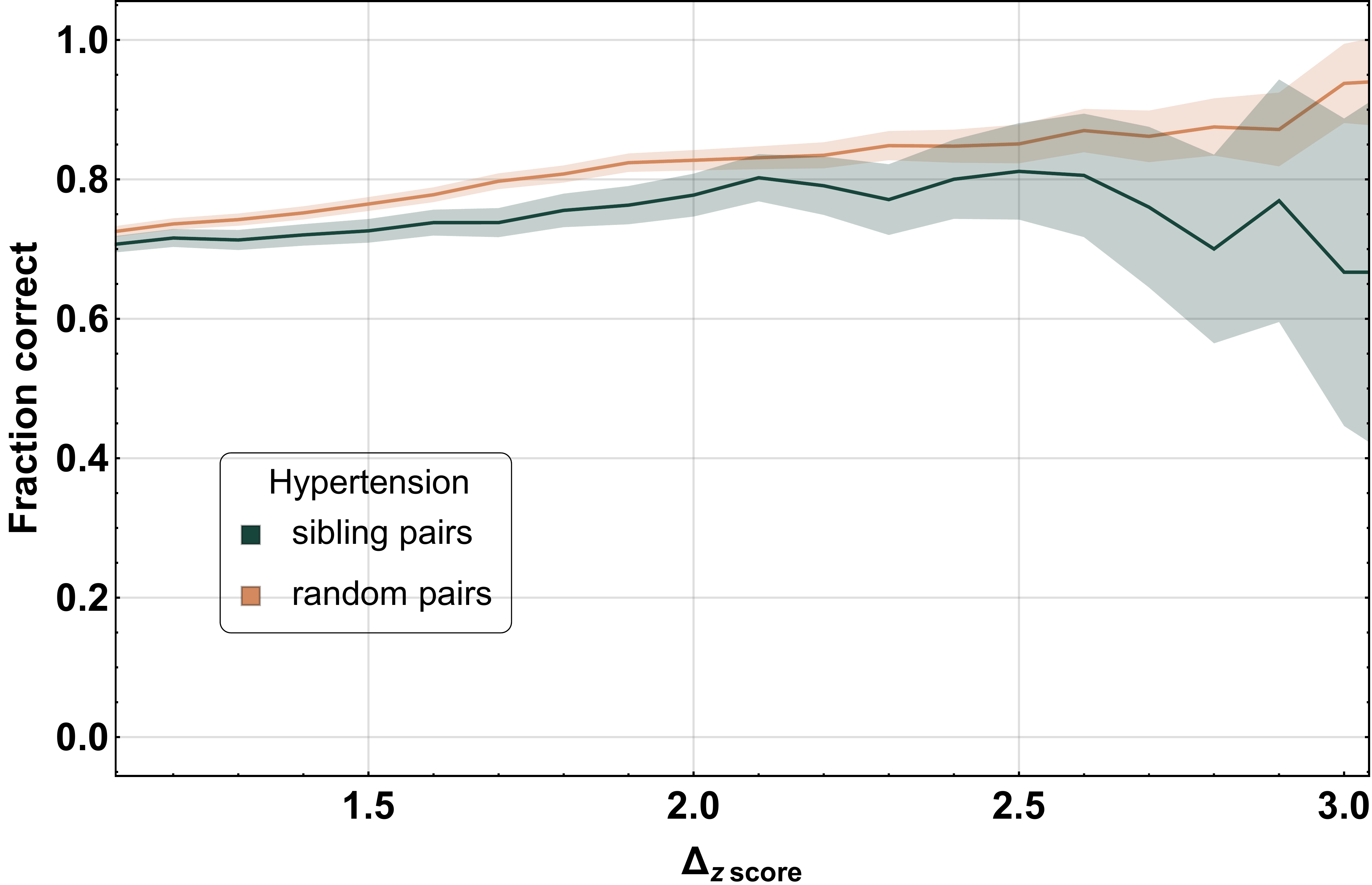} \\
\caption{
Predictors tested on random (non-sibling) pairs and affected sibling pairs with a single case, for conditions coronary artery disease (CAD), type 1 diabetes (T1D), hypothyroidism, and hypertension. One individual in each pair is high risk (i.e., has a high polygenic risk score) and the other is normal risk ($\text{PRS} < + 1 \text{SD}$).
The difference in PRS z-scores is given on the horizontal axis. The individual classified as high risk by the predictor is likely to be the one to exhibit the condition, increasingly so as the z-score difference becomes large. Quality of prediction is very similar between pairs of random individuals and sibling pairs, despite the siblings having experienced more similar childhood environments and sharing more alleles in common. Error bands include uncertainty due to limited numbers of individuals in each z-score bin. Reproduced from \cite{Lello:2020aa}.
}
\label{fig:frac}
\end{figure}

\begin{figure}[h!]
\includegraphics[width=0.95\linewidth]{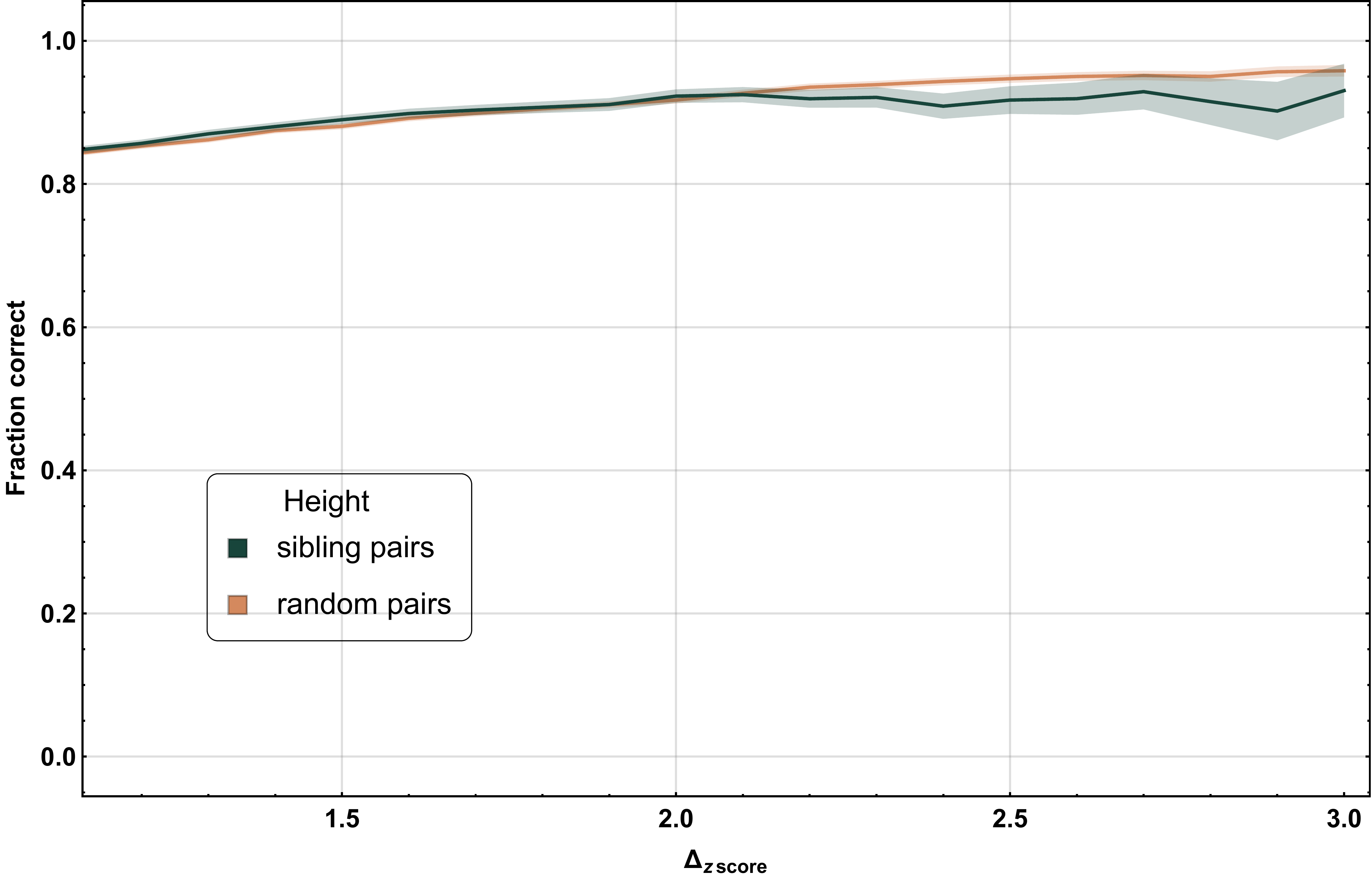}
\caption{
Probability that polygenic predictor correctly identifies the taller individual (vertical axis) for pairs of random individuals and pairs of siblings. Horizontal axis shows absolute difference in z-scored height between the individuals in each pair. Quality of prediction is very similar between pairs of random individuals and sibling pairs, despite the siblings having experienced more similar childhood environments and sharing more alleles in common. Error bands include uncertainty due to limited numbers of individuals in each z-score bin. Reproduced from \cite{Lello:2020aa}.
}
\label{fig:heightfrac}
\end{figure}

We tested a variety of polygenic predictors using tens of thousands of genetic siblings for whom we have SNP genotypes, health status, and phenotype information in late adulthood. Siblings have typically experienced similar environments during childhood, and exhibit negligible population stratification relative to each other. Therefore, the ability to predict differences in disease risk or complex trait values between siblings is a strong test of genomic prediction in humans. We compare validation results obtained using non-sibling subjects to those obtained among siblings and find that typically most of the predictive power persists in within-family designs. Given 1 sibling with normal-range PRS score (less than 84th percentile) and 1 sibling with high PRS score (top few percentiles), the predictors identify the affected sibling about 70-90 percent of the time across a variety of disease conditions, including breast cancer, heart attack, diabetes, etc. For height, the predictor correctly identifies the taller sibling roughly 80 percent of the time when the (male) height difference is 2 inches or more. Some disease prediction results are illustrated in \cref{fig:frac} while a sibling validation of height prediction can be seen in \cref{fig:heightfrac}.

\section{Methods, Genetic Architecture, and Theoretical Considerations} 

\subsection{Sparse learning, L1 penalization, phase transition behavior}
Sparse learning algorithms have been successfully applied to construct genomic predictors \cite{compressed,height,Vattikuti2014,lee2016uncovering}. These algorithms incorporate a prior that SNPs which materially affect the trait are only a small fraction of the (typically of order million or more) candidate SNPs. In other words, the algorithms favor parsimony in the construction of models for genetic risk. This prior has been confirmed by many studies: most of the population variance for even the most polygenic traits (e.g., human height) is captured by at most tens of thousands of SNPs \cite{lello2018accurate}. Although 10k is a large number, it is small compared to the millions of candidate common SNPs (i.e., polymorphisms found in at least $\sim 1$ percent of the population) in each person's DNA. Hence, the assumption of sparsity has strong empirical support \cite{lambert2020polygenic}.

Our lab has successfully used a sparse learning technique called L1 penalized regression, also known as LASSO or compressed sensing. Below we elaborate on our methodology in detail.

In analyses performed to date, LASSO, and penalized regression in general, performs as well or better than other training algorithms \cite{prive2019efficient} --- such as logistic regression, support vector machines, linear mixed-models, random forests, Bayesian regression \cite{Wei_2009,Abraham2012,Botta_2014,okser2014regularized,delosCampos2015,carvalho2010horseshoe,chang2019probabilistically,berger2015effectiveness} --- for trait prediction. Studies comparing sparse learning against other methods, including neural networks, have not found a consistent advantage; it is fair to say that currently sparse learning methods are comparable to or better than the alternatives \cite{bellot2018can,azodi2019benchmarking,gustavo1}.

\begin{figure}[h!]
\includegraphics[width=0.95\linewidth]{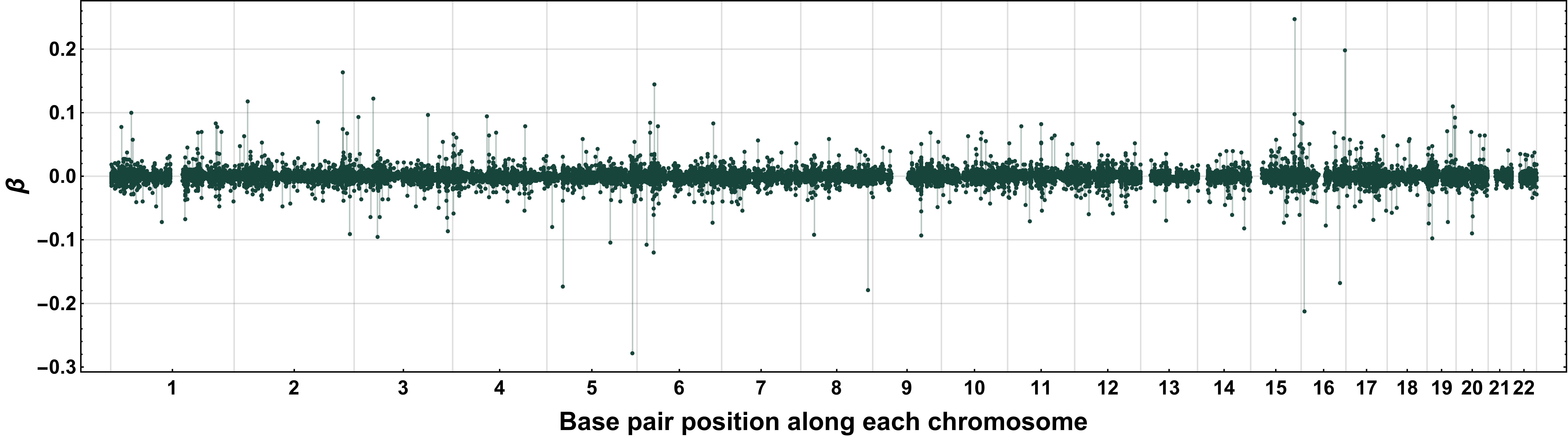}
\caption{
Locations of $\sim 20$k SNPs activated in height predictor on the genome, with individual chromosomes indicated. Vertical axis is effect size $\beta$ of minor allele. Positive $\beta$ value indicates a SNP for which the minor allele is associated with increased height. Reproduced from \cite{height}.
}
\label{fig:betas}
\end{figure}

Our standard process for building a sparse predictor was designed to optimize performance \emph{within} an specific ancestry group, as self-reported or according to PCA clustering. We elaborate more on cross-ancestral studies later in this section while for this method, both training and validation is performed within a specified ancestry.
\begin{description}
\item[Predictor training] In order to avoid difficulties arising from population structure, training is performed in a homogeneous population with similar ancestry. Standard tools using, e.g., principal components analysis, allow efficient categorization by ancestry. The set of candidate SNPs is typically either the full set of SNPs directly measured by the genotyping array, or a larger set obtained by imputation. Basic quality control is performed to avoid using extremely rare variants and poorly genotyped participants. The weights ${\beta_j}$ are chosen by minimizing the objective function 
$$
O = \tfrac{1}{2} \vert \vec{y} - X\vec{\beta} \vert^2 + \lambda \sum_j |\beta_j|
$$
for the vector of phenotypes $y$ and matrix $X$ of SNP genotypes for each sample. $\lambda$ is a hyperparameter of the model which tunes the level of sparsity imposed. The predictors are trained using $k$-fold cross-validation: a small subset of data is withheld from training and used for model selection; the entire process is repeated $k$ times with a different subset withheld every time.

\item[Score/Validation] Each trained predictor is scored on its corresponding validation subset withheld from its $k$-fold training. We typically use standard prediction metrics such as AUC and explained variance to determine optimal parameter settings and to select top predictors from each cross-validation fold.

\item[Evaluation] Once the optimal predictors have been selected they can be evaluated in a number of ways. Typically we use evaluation data sets composed of \initem{1} individuals of similar ancestry to the training set, but not used in training, \initem{2} individuals of adjacent ancestry not used in training (e.g., Eastern or Southern Europeans, adjacent to British / North-West Europeans), \initem{3} individuals of generally similar ancestry but collected in entirely different cohorts, sometimes from another continent, \initem{4} distant ancestry groups (e.g., non-Europeans of a specific ancestry such as East Asian or African), and most prominently \initem{5} \emph{siblings} of generally European ancestry (see \cite{Lello:2020aa} for a full paper on sibling evaluation). We again compute standard prediction metrics such as AUC, but also more relevant measures such as the absolute probability (rate of incidence) of the condition in outlier subgroups such as top few percent PRS score, as illustrated in \cref{fig:prevalence}.
\end{description}
Predictors built with L1 penalization typically have between 100 and 20k active SNPs, depending on phenotype, distributed over many chromosomes --- see \cref{fig:betas} for the example of height. The predictor performance varies with the trait but is always strongly dependent on the sample size available for training. Empirically, we find an approximate sample size dependence $\sim \tfrac{N}{N+b}$ with the asymptotic behavior determined by the (linear/narrow-sense) heritability of the trait in question and by the number of candidate SNPs. In practice, there is typically a very steep gain in prediction power as the sample sizes grow from small to moderate, say from 1k to 50k samples. It is then followed by a region of less dramatic but steady performance gains until it eventually flattens out for extremely large data sets. \Cref{fig:auc} shows how predictor performance improves with increasing sample sizes for hypertension, hyperthyroidism and type 2 diabetes.

We have shown, e.g., for height and bone-heel mineral density, that the UKB with its $\sim$ 400k individuals of European ancestry is currently in the late second region. For many traits there is still substantial predictive power to be gained from increased sample size, as seen in \cref{fig:auc}. 

Sparse learning algorithms using L1 penalization have been demonstrated to be highly effective. Our original interest in these methods arose because, almost uniquely, there are strong mathematical results characterizing their behavior. For example, celebrated theorems (largely unknown to the computational genomics community) provide performance guarantees for compressed sensing \cite{Vattikuti2014} as a function of signal/noise ratio and training data size. Further, it is known that these algorithms display a kind of universal phase transition behavior as the data size is varied. In \cite{compressed} we showed that matrices of human SNP genotypes are good compressed sensors, and exhibit the celebrated Donoho-Tanner phase transition behavior previously found for large classes of random matrices. These results were further verified in \cite{de2018complex,height}.

\subsection{Prediction across ancestry groups, causal variants}
PRS training has so far overwhelmingly been conducted in populations of homogeneous ancestry, typically of European descent, for which large databanks first became available. This is because population stratification (patterns of correlation within the genome, which differ by ancestry) introduces special difficulties in statistical learning (e.g. \cite{sohail2019polygenic}). Consequently, the majority of predictors have been trained on and work best in European populations. There are a few exceptions in which GWAS works well in diverse populations, e.g. \cite{loos2014bigger}, but performance of complex PRS fall off quickly as a function of genetic distance \cite{10.1371/journal.pbio.1001661,martin2017human}. The implications of this skewed focus are serious as the majority of the world population, including minorities within countries of predominantly European descent, are left out from these new advancements in health care \cite{Martin441261}. It is thus an urgent priority to correct this situation: \initem{1} by building predictors using cohorts from other ancestries (e.g., East Asian or West African) as well as \initem{2} developing techniques that can modify or adapt predictors trained in one ancestry group so that they work well in another, perhaps distant, ancestry group.

Cross-ancestral study and training of predictors provide a unique opportunity to explore the genetic architecture for common diseases, a research area in which important basic questions still remain. While PRS and GWAS identify candidate SNPs which are statistically associated with increased (or decreased) risk, they cannot determine which SNPs have a causal effect on individual biology. Because SNPs often occur in correlated clusters in the genomes of a given population, there is always some ambiguity concerning whether a specific SNP is causal. These correlation patterns vary across populations.

Current predictors utilize SNPs which are merely {\it tags} (i.e., correlated in state) for the {\it actual causal} SNPs (or other structure). The quality of the tag may be much weaker in a distant population, causing the predictor to perform much worse. This is a problem to be solved.

By utilizing different patterns of correlation, we may be able to {\it zero in on} actual causal SNPs --- they are likely to be {\it consistently detected} in predictor training across multiple distant populations. In other words, we can turn the problem described above into a tool for detecting candidate {\it causal} SNPs. See for instance \cite{Koyama_2020} for early results for CAD and \cite{Dehghani2020.06.11.146993} for a current overview of how cross-population studies have furthered the understanding of Alzheimer's Disease.

\subsection{Coding regions, Pleiotropy}
Using the SNPs activated in existing PRS, we can begin to investigate the genetic architecture of common disease risk \cite{yong}. There are many detailed results concerning specific conditions, but two general points should be emphasized:
\begin{enumerate}
    \item Much of the genetic risk identified in polygenic predictors is controlled by variants outside genic (protein coding) regions, and not accessible through exome sequencing. This supports the notion that DNA information storage extends beyond specific genes. 
    \item The DNA regions used in disease risk predictors so far constructed seem to be largely disjoint, suggesting that most genetic disease risks are largely uncorrelated. It seems possible in theory for an individual to be a low-risk outlier in all conditions simultaneously.
\end{enumerate}
The space of genetic variation is high dimensional, and extends far beyond individual (protein coding) genes. Intuitions about strong pleiotropy are likely wrong --- they were developed before we knew anything about real genetic architectures. There seem to be many causal variants that could, in principle, be independently modified and evidence to date suggests that large portions of genetic variance affecting different human traits and disease risks are independent. 

In the final section below, we make some rough estimates concerning the total space of heritable individual differences (including both quantitative traits and disease risks) for humans, assuming approximate independence.

\subsection{Linearity / Additivity}
There is significant empirical evidence now that linear predictive models can capture much (nearly all?) of the estimated common SNP heritability of many traits. It may come as a surprise that genetic effects can be approximately additive, given the apparent complexity of biological systems.

Nonlinear genetic effects certainly exist and are likely realized in
every organism. However, quantitative differences between individuals within a species may
be largely due to independent linear effects of specific genetic variants. As reflected in Fisher's Fundamental Theorem of Natural Selection \cite{Ho2015}, linear
effects are the most readily evolvable in response to selection, whereas nonlinear ``gadgets'' (i.e., mechanisms which depend sensitively on multiple genetic switches) are
more likely to be fragile to small changes. Evolutionary adaptations requiring significant
changes to nonlinear gadgets are improbable and therefore require exponentially more time
than simple adjustment of frequencies of alleles of linear effect. One might say that to first
approximation, Biology = linear combinations of nonlinear gadgets, and most of the variation
between individuals in a species is due to the (linear) way gadgets are combined, rather than in the realization
of different gadgets in different individuals.

\bigskip

\section{In Vitro Fertilization and Genetic Engineering}

Today millions of babies are produced through In Vitro Fertilization (IVF). In most developed countries roughly 3-5 percent of all births are through IVF, and in Denmark the fraction is about 10 percent \cite{DenmarkARTStat2018}. But when the technology was first introduced with the birth of Louise Brown in 1978, the pioneering scientists had to overcome significant resistance.
\begin{quote}
Wikipedia: ...During these controversial early years of IVF, Fishel and his colleagues received extensive opposition from critics both outside of and within the medical and scientific communities, including a civil writ for murder. Fishel has since stated that "the whole establishment was outraged" by their early work and that people thought that he was "potentially a mad scientist". \cite{wiki:Fishel}
\end{quote}
In the past, parents with more viable embryos than they intended to use made a selection based on very little information --- typically nothing more than the appearance or morphology of each blastocyst. With modern technology it has become common to genotype embryos before selection, in order to detect potential genetic issues such as trisomy 21 (Down Syndrome).  Parents who are carriers of a single gene variant linked to a Mendelian condition can use genetic screening to avoid passing the risk variant on to their child. Millions of embryos are now genetically tested each year. With polygenic risk prediction, it is possible now to screen against outlier risk for many common disease conditions, not just rare single gene conditions. For example, the overwhelming majority of families with breast cancer history are not carriers of a BRCA risk variant, but rather have elevated polygenic risk. It is now possible for these families to select an embryo with average or even below average breast cancer risk if they so wish. 

Beyond IVF embryo selection, the advent of CRISPR and other recent advances in genetic editing suggest that future technologies will permit germ line editing of humans --- perhaps leading to consequences in human evolution. 

Note that for genomic prediction it is enough to identify SNPs which are correlated to the phenotype. But to achieve the desired effect in editing one must identify the actual {\it causal} genetic variants. This step in the research program is highly nontrivial and may take longer to accomplish than the development of the molecular editing tools.  

Highly polygenic traits imply a very large reservoir of extant variance already present in the population \cite{Ho2015}. It is this extant variance that plant and animal breeders have used to advance agriculture for thousands of years. Roughly speaking, if a trait is controlled by $\sim N$ genetic variants, an increase in phenotype by one population standard deviation corresponds to changing $\sim N^{1/2}$ variants from the $(-)$ to $(+)$ state. Thus the maximum number of standard deviations that can be captured through editing could be as large as $N^{1/2}$!

The population geneticist James Crow of Wisconsin wrote \cite{Crow2010}:
\begin{quote}
The most extensive selection experiment, at least the one that has continued for
the longest time, is the selection for oil and protein content in maize (Dudley
2007). These experiments began near the end of the nineteenth century and still
continue; there are now more than 100 generations of selection. Remarkably,
selection for high oil content and similarly, but less strikingly, selection for high
protein, continue to make progress. There seems to be no diminishing of selectable
variance in the population. The effect of selection is enormous: the difference in
oil content between the high and low selected strains is some 32 times the original
standard deviation.
\end{quote}
To take another example, wild chickens lay eggs at the rate of roughly one per month. Domesticated chickens have been bred to lay almost one egg per day. (Those are the eggs we have for breakfast!) Of all the wild chickens in evolutionary history, probably not a single one produced eggs at the rate of a modern farm chicken.

The corresponding ethical issues are complex and wide ranging. They deserve serious attention in what may be a relatively short interval before these capabilities become a reality in human genetic engineering and widespread clinical practice. We cannot do them justice here, but they include topics such as: the power dynamic between population geneticists and studied populations \cite{brodwin2005bioethics}; personal and communal identity issues \cite{juengst2004face}; comparisons of different types of pre-implantation testing \cite{braude2002preimplantation,geraedts2009preimplantation}; how to develop ethical guidelines in a fast changing field \cite{londra2014assisted,treff2019utility}; disease specific concerns \cite{sabatello2020ethics}; misinformation and media bias \cite{venturella2017modern}; congenital vs adult-onset testing \cite{ethics2018use}; non-medical testing and sex selection \cite{ethics2015use,ethics2018disclosure}; provider duties and obligations \cite{ethics2017transferring}; disparities in health care \cite{martin2017human}; intersection of legal and religious concerns with genetics \cite{zuradzki2014situation,sandel2004embryo}; evolution of the limits of gene editing \cite{brokowski2019crispr}; ethical statements and disclosures of those using CRISPR \cite{brokowski2018crispr}; patent competition and human application of CRISPR \cite{peng2016morality}; animal welfare \cite{schultz2018crispr}; and the extensive and intertwining history of genetics and eugenics \cite{schulman1996preimplantation, levine2010introduction, ekberg2007old, wikler1999can}. 

Each society will decide for itself where to draw the line on human genetic engineering, but we can expect a diversity of perspectives. Almost certainly, some countries will allow genetic engineering, thereby opening the door for global elites who can afford to travel for access to reproductive technology. As with most technologies, such as IVF, the rich and powerful will be the first beneficiaries. Eventually, though, it is possible that many countries will not only legalize human genetic engineering, but even make it a (voluntary) part of their national healthcare systems. The alternative would be inequality of a kind never before experienced in human history.

\section{The Future}

We conclude with some predictions concerning future progress, and some unifying theoretical remarks related to high dimensionality and its role in genomic prediction.

Perhaps the easiest prediction to make is that there are still significant gains to be made from simply increasing the training data size. Already for some common disease conditions we can identify outliers (e.g., few percent of the population) who have well over 50 percent probability to have the disease by late adulthood. This level of prediction will become available for many more conditions, and the size of the identifiable high risk population will increase considerably. There will be important consequences for early screening, diagnosis, and prevention in medical care. Health insurance may also be transformed: individuals who know their risk profiles have an asymmetric information advantage over insurers. Perhaps we will see a day when no insurer will price a policy without DNA analysis. Or, perhaps strong genomic prediction will force societies into a single payer healthcare structure, in which all risks are pooled.

Next we list some startling developments that can already be anticipated and only await sufficient training data to be realized. 
\begin{enumerate}
    \item Face Recognition: AI algorithms use of order 100 features to identify human faces (e.g., distance between the eyes, or from nose to mouth, etc.). From identical twins, we know that these features --- each a complex quantitative trait --- is itself highly heritable. With enough training data, already conveniently extracted by face recognition algorithms from ordinary photos, we expect that facial features will be predictable from DNA and hence faces themselves can be reconstructed from DNA alone. This will likely have applications in forensic science (e.g., to solve crimes) as well as in IVF (parents will have an idea of what their child will look like, at difference ages, from an embryo genotype). It will also intersect with the ethical use of facial recognition technology and the protection of civil liberties \cite{bowyer2019face, herschel2017ethics, brey2004ethical}. 
    \item Cognitive and Personality traits: Substantial progress has been made in the prediction of cognitive ability \cite{belsky2018genetic,Lee2018}. Actual cognitive score correlates roughly 0.3 to 0.4 with predicted score. Quality of prediction appears to be entirely data limited, and we expect these correlations to increase considerably before the regime of diminishing returns is reached. Personality traits such as conscientiousness, extraversion, or agreeableness are known to be highly heritable, and we expect them to be predictable from genotype once sufficient phenotype data become available \cite{Jang1996}. In general, behavioral traits are more greatly influenced by environmental factors than other phenotypes, exacerbating the above mentioned ethical concerns.
    \item Longevity: As we mentioned in our discussion of pleiotropy, genetic disease risks seem to be largely (although not completely) independent. This suggests that (at least theoretically), individuals could exist that have low risk across all common disease conditions that impact life expectancy. In the future we may be able to identify outliers in longevity, and understand better the limits to human life span.
\end{enumerate}
We can begin to formulate a ``grand unified theory'' of human individual differences, using an information theoretic approach and what we already know about the genetic architecture and dimensionality of the space of human variation \cite{Meisner2020}. Some orders of magnitude:
\begin{itemize}[nosep]
    \item $\sim$10M common SNP differences between two individuals.
    \item $\sim$10k SNPs may control most of the variance for a typical complex trait.
\end{itemize}
We can take the number of common (e.g., MAF $> 0.01$) SNP differences that is typical between two individuals as an estimate of the amount of genomic information that determines the (heritable) individual differences among humans. In principle, there could be $\sim$1k (a few thousand) largely independent complex traits with little pleiotropy between them. These might include a hundred common disease or health risks, hundreds of cosmetic traits, including facial and body morphology parameters, dozens of psychometric variables, including personality traits, etc. 

Clearly, individual differences are well accommodated by a $\sim$1k dimensional phenotype space embedded in a $\sim$10M dimensional space of genetic variants.

Of course, it is an unrealistic idealization for the traits to be entirely independent in genetic architecture. We expect to find pleiotropy at some level: a subset of genetic variants will affect more than one trait. But these estimates suggest that a significant part of the genetic variance of each trait can be predicted using existing linear models, and potentially modified (e.g., via editing) independently of the other traits. This is simply a consequence of high dimensionality.

It seems certain that genomic prediction will have significant impacts on health care through better screening, diagnosis, and treatment of almost all important disease conditions. As we have emphasized the main limiting factor is the availability of sufficiently large data sets with good phenotype information across populations of diverse ancestry. Computational and algorithmic methods are not (at least at the moment) the main constraint on progress. Perhaps some day soon we will have good predictors for almost all heritable individual differences in the human species. The dream of decoding the human genome is within our reach
and important societal choices concerning how to apply the results are upon us.

\paragraph{Competing Interests}
Stephen Hsu a shareholder of Genomic Prediction, Inc. (GP), and serves on its Board of Directors. Louis Lello is an employee and shareholder of GP. Timothy Raben and Erik Widen have no commercial interests relevant to this research.

\newpage
\FloatBarrier
\printbibliography
\end{document}